\documentclass[conference,8pt]{article}

\usepackage{spconf,amsmath,graphicx}

\title{Enhancing Age-Related Robustness in Children Speaker Verification}
\name{Vishwas M. Shetty\textsuperscript{1,*},
Jiusi Zheng\textsuperscript{1,*},
Steven M. Lulich\textsuperscript{2},
Abeer Alwan\textsuperscript{1}\thanks{* These authors contributed equally to this work.}
}
\address{\textit{\textsuperscript{1}Department of Electrical and Computer Engineering, University of California, Los Angeles, USA}\\
\textit{\textsuperscript{2}Speech Language and Hearing Sciences, Indiana University, Bloomington, Indiana, USA}\\
\textit{\{shettyvishwas, zheng94\}@ucla.edu, slulich@iu.edu, alwan@ee.ucla.edu}}
\usepackage{cite}
\usepackage{amsmath,amssymb,amsfonts}
\usepackage{algorithmic}
\usepackage{graphicx}
\usepackage{textcomp}
\usepackage{adjustbox}
\usepackage{caption}
\usepackage{comment}
\usepackage{multirow}
\usepackage{url}
\usepackage{float}
\usepackage{tablefootnote}

\usepackage{xcolor}
\def\BibTeX{{\rm B\kern-.05em{\sc i\kern-.025em b}\kern-.08em
    T\kern-.1667em\lower.7ex\hbox{E}\kern-.125emX}}

\begin{document}
\fontsize{9.5pt}{11pt}\selectfont
\maketitle
\begin{abstract}
One of the main challenges in children's speaker verification (C-SV) is the significant change in children’s voices as they grow. In this paper, we propose two approaches to improve age-related robustness in C-SV. We first introduce a Feature Transform Adapter (FTA) module that integrates
local patterns into higher-level global representations, reducing overfitting to specific local features and improving the inter-year SV performance of the system. We then employ Synthetic Audio Augmentation (SAA) to increase data diversity and size, thereby improving robustness against age-related changes. Since the lack of longitudinal speech datasets makes it difficult to measure age-related robustness of C-SV systems, we introduce a longitudinal dataset to assess inter-year verification robustness of C-SV systems. By integrating both of our proposed methods, the average equal error rate was reduced by 19.4\%, 13.0\%, and 6.1\% in the one-year, two-year, and three-year gap inter-year evaluation sets, respectively, compared to the baseline.
\end{abstract}

\section{Introduction}
\label{sec:intro}
With the advent of online learning, the number of children using online learning applications has increased exponentially. Speaker verification can enable personalized experiences for children in educational applications, virtual assistants, and interactive learning tools by recognizing individual voices. This can enhance engagement and tailor learning materials to each child. The current state of the art speaker verification systems \cite{ECAPA_TDNN, VILLALBA2020101026, zhang2022mfa, xiang2019margin} have achieved significant success on various benchmark datasets\cite{nagrani2017voxceleb,hechmi2021voxceleb, cole1998cslu, zheng20233d}. While speaker verification for adults has seen significant improvements in accuracy due to large datasets and optimized model architectures, speaker verification for children faces greater difficulties.

The challenges in children speaker verification are more pronounced compared to adults in the following aspects. First, the limited availability of children’s speech datasets makes it difficult for models to reach their full potential, particularly because insufficient data can lead to overfitting, which in turn reduces the model's robustness. Many studies have addressed this issue in children's speaker verification by employing various data augmentation techniques, including SpecAug \cite{park2019specaugment}, Speed Perturbation \cite{ko2015audio}, Noise and RIR \cite{ko2017study,snyder2015musan}, Voice Conversion \cite{shahnawazuddin2020domain,aziz2024enhancing,aziz2024short}, and a combination of several signal processing approaches \cite{singh2024childaugment}. Second, the anatomical changes in children's vocal tracts during growth lead to more pronounced variations in their voices \cite{lee1999acoustics, yeung2018difficulties, shetty2023developmental}, compared to the relatively stable vocal characteristics of adults \cite{santos2023vocal}.
Due to the scarcity of longitudinal speech datasets, research on the impact of age-related changes on the performance of speaker verification systems is limited \cite{qin2022cross}, with a few studies addressing inter-year C-SV. However, understanding and improving age robustness in children's speaker verification systems is crucial in practical applications because children's voices can change significantly over short periods, leading to potential degradation in verification accuracy over time. Addressing this issue is essential to ensure consistent and reliable performance of these systems in real-world scenarios where a child may use the verification system over several years. 


In this paper, we propose techniques to improve the inter-age C-SV through two approaches. Drawing inspiration from recent advancements in feature adapters \cite{tian2024learning, huang2024robust}, we first propose the Feature Transform Adapter (FTA) which aggregates the original local features at different time scales and resolutions to eliminate disturbances caused by age-related changes in the speaker's audio characteristics, thereby forming more robust features. We then present a data augmentation method using HiFi-GAN \cite{kong2020hifi} to generate in-domain synthetic audio from speech spectrograms. Although this approach has been previously applied in other fields \cite{rossenbach2020generating, madhu2019data}, we have adapted this data augmentation technique for the children SV task. As a final step, we introduce a longitudinal children's speech dataset designed to support the evaluation of age-related speech changes and to advance the development and testing of robust speaker processing systems for children. A comprehensive set of ablation studies is conducted to validate the efficacy of the proposed methods.

\section{Method}
\label{sec:Methods}

\subsection{Feature Transform Adapter}
\label{subsec:feature_transform_adapter}

\begin{figure}[t]
    \centering
    \scalebox{0.50}{  
        \includegraphics[width=1.0\columnwidth]{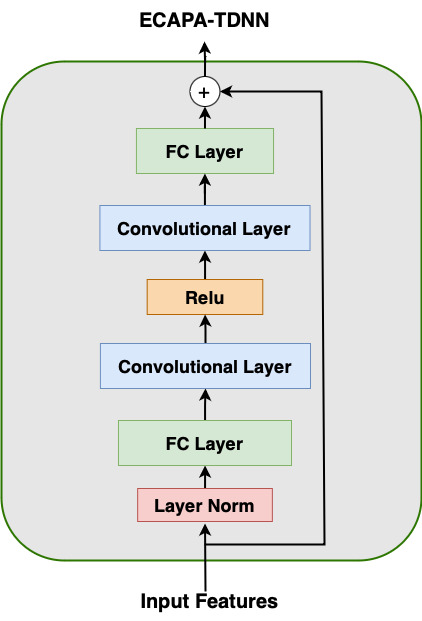}  
    }
    \caption{Overview of the proposed Feature Transform Adapter}
    \label{fig:Proposed_framework}
\end{figure}

Different from inter-year speaker verification for adult speech \cite{qin2022cross} (referred to as cross-age in \cite{qin2022cross}), age significantly impacts the performance of C-SV systems. In this context, global features, such as formant distribution\cite{lulich2020development}, refer to the stable and consistent aspects of a speaker’s voice that remain relatively unchanged over time, making them crucial for identifying the speaker’s identity. In contrast, local features, such as segmental durations\cite{lee1999acoustics}, exhibit transient, short-term variations that are sensitive to slight changes in the input signal, such as age-related shifts. We hypothesize that as a child ages, local features undergo changes in both the time and frequency domains, whereas global features remain relatively stable. To address this, we introduce an innovative structure referred to as the Feature Transform Adapter (FTA), as depicted in Fig \ref{fig:Proposed_framework}. This adapter aggregates features in a manner that obscures age-related local information while preserving essential global information related to the speaker’s identity. The primary goal of this adapter is to enhance the robustness of the children's speaker verification system, making it more effective in handling age-related variability and improving overall verification accuracy.

The details of the FTA are as follows. Initially, the 80-dimensional filterbank features are layer-normalized and passed through a fully connected layer. The features are then processed through two 1D convolutional layers to aggregate local features along the frequency dimension. The aim is to capture the average variations along the frequency dimension while maintaining the global speaker identity information. The ReLU activation function is employed between the two convolutional layers to introduce non-linearity into the adapter. Finally, the processed features are integrated with the original features via a residual connection. These features can be then used to train the SV system. 

\subsection{Synthetic Audio Augmentation}
\label{subsec:Synthetic_Audio_augmentation}
Synthetic Audio Augmentation (SAA) involves using artificially generated audio samples to diversify training datasets. We assume that such synthesis methods preserve the speaker’s identity but may not accurately retain age-related information due to challenges in capturing age-specific features during synthesis. In this paper, we introduce an SAA method based on HiFi-GAN \cite{kong2020hifi}. 
HiFi-GAN includes one generator and two discriminators: the generator converts mel spectrograms into raw waveforms through the convolutional neural network. Among the discriminators, the Multi-Period Discriminator (MPD) captures diverse periodic patterns by analyzing various segments of the audio, while the Multi-Scale Discriminator (MSD) focuses on both short-term and long-term dependencies across multiple resolutions. During training, the generator learns to produce realistic audio waveforms that can deceive both discriminators, while the discriminators simultaneously refine their ability to distinguish between real and generated audio. In this work we directly use the pre-trained open source HiFi-GAN generator \cite{hifigan_github} to generate the synthetic audio. For each child’s speech signal from a specific year, we input the corresponding spectrogram into the HiFi-GAN generator, generating synthetic audio to augment the training dataset.

\subsection{Child Longitudinal Test Set}
\label{subsec:Child_Longitudinal}

To address the dataset gap in inter-year children speaker verification, we present a longitudinal evaluation set. Derived from an existing longitudinal child speech dataset \cite{shetty2023developmental, charles2019ultrasound, lulich2020development} and referred to as the IU dataset, it includes data from eight children (four boys, four girls) recorded annually from Grade 1 to Grade 4. The speech utterances were elicited through tasks commonly used by speech-language pathologists (Goldman and Fristoe Test of Articulation, 3rd Edition, aka GFTA-3) and were recorded at Indiana University, Bloomington.  We applied Voice Activity Detection (VAD) \cite{Silero_VAD} to extract child speech segments of at least 2 seconds, creating the IU evaluation database.

IU inter-year verification pairs were created by pairing enrollment speech from one grade with test segments from subsequent grades. For example, Grade 1 (G1) enrollment speech was paired with Grade 2 (G2), Grade 3 (G3), and Grade 4 (G4) segments, forming G1-G2, G1-G3, and G1-G4 target pairs. Negative pairs were sourced from the same year to increase task complexity. Other sets like G2-G3, G2-G4, and G3-G4 were constructed similarly. Additionally, IU intra-year sets, i.e., G1-G1, G2-G2, G3-G3, and G4-G4 were created. Each IU evaluation set contains 2,000 verification pairs—1,000 positive and 1,000 negative.

\section{Experimental Settings}
\label{sec:Experimental_Settings}

\subsection{Databases}
\label{subsec:Database}
The CSLU \cite{shobaki2000ogi} dataset is utilized in all the training and fine-tuning experiments proposed in this paper. The CSLU  speech corpus, also referred to as the OGI kids' speech corpus, includes both spontaneous and scripted speech data from approximately 1,100 children, ranging from kindergarten (\textit{K00}) through grade 10 (\textit{K10}). Each age group consists of approximately 100 speakers recorded uttering single words, sentences, and digit sequences. For our experiments, we utilized only the scripted speech data.

We employ two distinct train-eval splits from the CSLU database, as detailed in Table 1. To achieve optimal results through fine-tuning and our proposed methods, we use CSLU-S, a split with a large training set, providing ample data for fine-tuning. As a result, the evaluation set in CSLU-S contains fewer speakers, with 866 speakers in the training set and 255 speakers in the evaluation set, covering 44,000 intra-year SV trials across all grades. To validate the effectiveness of our methods with a larger number of speakers in intra-year SV and to compare our approaches with existing state of the art C-SV models, we utilize the CSLU-L dataset in some of the experiments. In this split, the training set contains 120 speakers, and the evaluation set includes 993 speakers with 190,972 SV trials. There is no overlap between the speakers in the train and evaluation subsets of both splits. Along with the evaluation sets from CSLU database, we also evaluate the models on IU dataset discussed in section \ref{subsec:Child_Longitudinal}.

\begin{table}[t!]
\centering
\caption{\textit{Train-eval splits used to train/fine-tune the models discussed in this paper. \#Spks refers to number of speakers, \#Hrs represents duration in hours, and \#Trials stands for number of SV trials in the evaluation set.}}
\resizebox{0.9\columnwidth}{!}{
\begin{tabular}{lcccc}
\hline 
\multirow{2}{*}{} & \multicolumn{2}{c}{Train} & \multicolumn{2}{c}{Eval} \\ \cline{2-3} \cline{4-5} 
                  & \# Spks  & \# Hrs  & \# Spks  & \# Trials  \\ \hline
\textit{CSLU-S \cite{fan2024benchmarking}}   & 866      & 24.00      & 255      & 44000      \\ 
\textit{CSLU-L \cite{singh2024childaugment}}   & 120      & 2.65     & 993      & 190972     \\ \hline
\end{tabular}
}
\label{tab:CSLU_database_split}
\end{table}



\subsection{Baseline System}
\label{subsec:baseline_description}
An ECAPA-TDNN \cite{ECAPA_TDNN} network was trained for a speaker identification (SID) task using the CSLU dataset, with the training process conducted via the SpeechBrain toolkit \cite{SpeechBrain}. The input features consisted of 80-dimensional filter bank features, extracted every 10 ms using a 25 ms window. 
The ECAPA-TDNN architecture includes frame-level convolutional layers with 1024 channels and 128 dimensional attention channels. The output of the model is the 192 dimensional speaker vectors. We then compare these vectors to examine speaker similarity. In terms of the training details, all models are trained for 15 epochs with a batch size of 16. The Adam optimizer is utilized with a learning rate of $1e{-3}$. Additionally, four online data augmentation methods were employed during training, with detailed parameter configurations provided in \ref{subsec:Augmentation}. This model is used as the baseline in our experiments and is referred to as \textit{Baseline} in this paper.

\vspace{-0.5mm}
\subsection{Augmentation Setup}
\label{subsec:Augmentation}
To construct a strong baseline system, we incorporated four distinct data augmentation techniques, each with the following parameter configurations. For Noise Augmentation, the signal-to-noise ratio (SNR) of the added noise was varied between 0 and 15 dB. Additionally, Room Impulse Response (RIR) augmentation was employed to enhance the system's robustness in reverberant environments. The frequency drop augmentation allowed for the random removal of 1 to 3 frequency bands at a time, across the full frequency range, with a frequency band width of 0.05. In time drop augmentation, audio chunks ranging from 1000 to 2000 samples were randomly removed, with between 1 and 5 chunks being dropped per audio sample.  The baseline model described in \ref{subsec:baseline_description} was trained using the above discussed augmentation techniques on the CSLU train data. In the Synthetic Audio Augmentation process  discussed in Section \ref{subsec:Synthetic_Audio_augmentation}, we utilized the open-sourced pretrained HiFi-GAN vocoder and denoiser \cite{hifigan_github}, where the denoising strength was set to 0.005. 

\renewcommand{\arraystretch}{1.1} 
\begin{table*}[t!]
\caption{\textit{Equal Error Rate (EER) results for various techniques across different grades (K00–K10 and G1–G4, representing different age groups). 'Intra-Year' refers to enrollment and test segments from the same grade, while 'Inter-Year' indicates segments from different grades (e.g., G1-G2). Techniques evaluated include the baseline model, fine-tuning, FTA (Feature Transform Adapter), RA (Residual Adapter), and SAA (Synthetic Audio Augmentation), all using the 'CSLU-S' dataset. The lower EER on the CSLU test set compared to the IU test set for the same age group is likely due to CSLU's longer average audio duration, which provides more detailed speaker information.} } 
\centering
\resizebox{0.9\textwidth}{!}{%
\begin{tabular}{l c c c c c c c c c c c}
\hline 
\multicolumn{1}{l}{} & \multicolumn{11}{c}{\textit{CSLU Intra-Year}} \\ \cline{2-12} 
 & K00 & K01 & K02 & K03 & K04 & K05 & K06 & K07 & K08 & K09 & K10 \\ \hline
\textit{Baseline} & 13.2 & 12.85 & 10.45 & 9 & 9.55 & 7.05 & 4.7 & 4.9 & 4.5 & 3.45 & 4.4 \\ 
\textit{Finetune} & \textbf{9.55} & 10.65 & 9.05 & 6.95 & 7.25 & \textbf{4.1} & 4.05 & 2.95 & 2.75 & 1.85 & 3.3 \\ 
\textit{ + RA } & 9.7 & 11.5 & \textbf{8.75} & 7.5 & 7.1 & 4.35 & 3.85 & 3.2 & 2.6 & 1.95 & 3.45 \\ 
\textit{ + FTA} & 10.5 & \textbf{10.35} & 9.05 & 6.85 & 5.8 & 4.2 & 3.45 & \textbf{2.85} & \textbf{2.55} & \textbf{1.7} & \textbf{3.25}\\
\textit{ ++ SAA} & 10.6 & 12.1 & 9.65 & \textbf{6.8} & \textbf{5.75} & 4.4 & \textbf{3} & \textbf{2.85} & 2.6 & 2.65 & 3.6 \\ \hline
\multicolumn{1}{l}{} & \multicolumn{4}{c}{\textit{IU Intra-Year}} & \multicolumn{7}{c}{\textit{IU Inter-Year}} \\ \cline{2-5} \cline{7-12}
& G1-G1 & G2-G2 & G3-G3 & G4-G4 &  & G1-G2 & G1-G3 & G1-G4 & G2-G3 & G2-G4 & G3-G4 \\ \hline
\textit{Baseline} & 22.5 & 15.5 & 17.5 & 16.2 &  & 29.5 & 32.8 & 34.7 & 39.5 & 40.3 & 23.4 \\
\textit{Finetune} & 19 & 13.9 & 15.8 & 11 &  & 28.7 & 36.1 & 36.7 & 37.1 & 39 & 24.1 \\
\textit{ + RA } & 19.4 & 16.7 & 13 & 13.1 &  & 27.9 & 38 & 37.1 & 39.5 & 38.3 & 22.3 \\ 
\textit{ + FTA} & \textbf{16.4} & 12.9 & 12.2 & \textbf{10.6} & & \textbf{26.6} & 34.2 & 35.8 & 34.2 & 35.2 & 21 \\
\textit{ ++ SAA} & 19.9 & \textbf{12.2} & \textbf{8.1} & 13.1 & & 27.1 & \textbf{31.15} & \textbf{32.6} & \textbf{30.6} & \textbf{32.3} & \textbf{16.8} \\ \hline 
\end{tabular}}
\label{tab:CSLU_IU_Results}
\end{table*}

\section{Results and Discussion}
\label{sec:Results_Discussion}

\subsection{Performance of Feature Transform Adapter}
\label{subsec:Performance_of_FTA}
To evaluate the robustness of the proposed C-SV systems, Table \ref{tab:CSLU_IU_Results} shows the Equal Error Rates (EER) for different SV systems trained using CSLU-S and evaluated under intra-year and inter-year verification (i.e., longitudinal evaluation) on the CSLU and IU evaluation sets. Fine-tuning significantly reduces the EER for intra-year verification compared to the baseline but does not consistently improve inter-year performance, likely due to insufficient representation of inter-year scenarios during training, which leads to poor generalization and variable performance. The FTA approach mitigates this limitation by leveraging convolutional layers to capture generalizable patterns, resulting in improved results across most CSLU evaluation sets and all IU evaluation sets compared to the conventional fine-tuning. In contrast, replacing the convolutional layers with fully connected layers (\textit{+RA}) shows no improvement,  hence supporting our hypothesis on the ability of the convolutional layers to aggregate local features as a key to enhancing robustness.

\subsection{Effect of Synthetic Audio Augmentation}

Overall, using SAA to fine-tune along with \textit{FTA}, referred to as \textit{++SAA} in Table \ref{tab:CSLU_IU_Results}, further reduces the EER on IU inter-year evaluation sets compared to \textit{+FTA}, although SAA carries the risk of lowering intra-year verification accuracy. In inter-year verification, as shown in Table \ref{tab:CSLU_IU_Results}, SAA achieved the lowest EER in 5 out of 6 IU inter-year evaluation sets, demonstrating a clear improvement. Specifically, the average EER were reduced by 19.4\%, 13.0\%, and 6.1\% in the one-year, two-year, and three-year gap evaluation sets, respectively, compared to the baseline. In contrast, in intra-year verification, SAA resulted in a noticeable increase in verification EER in 7 out of 11 age groups on the CSLU dataset, and 2 out of 4 intra-year evaluation sets on the IU dataset also showed EER degradation compared to using only the FTA method. We noticed that this trade-off between inter-year and intra-year verification accuracy aligns with the previous study on inter-year adult speech verification \cite{qin2022cross} (inter-year is referred to as cross-age in \cite{qin2022cross}). 

\subsection{The Impact of Number of Speakers}
\label{subsec:Effect_of_Speaker_size}

To evaluate the generalizability of our proposed methods on a larger intra-year evaluation set, we retrained the models using the CSLU-L train-eval split, with results summarized in the first four rows of Table \ref{tab:Speaker_and_Feature_adapt_ablation}. Our methods demonstrate continued effectiveness despite the substantial increase in the number of speakers in the CSLU intra-year evaluation set. Specifically, the model fine-tuned with both SAA and FTA (i.e., \textit{++SAA\textsuperscript{\dag}}) reduced the EER by 28.3\% on the CSLU intra-year set, 29.3\% on the IU intra-year set, and 10.8\% on the IU inter-year set compared to the \textit{Baseline\textsuperscript{\dag}} model. In the CSLU-L case, fine-tuning with only the FTA adapter (\textit{+FTA\textsuperscript{\dag}}) showed no improvements. One possible explanation is that the CSLU-L train set has a significantly lower number of speakers and hence the limited size of the training set becomes the dominant factor constraining model performance. This hypothesis is supported by the considerable performance gains from incorporating the SAA data augmentation method during fine-tuning (i.e., \textit{++SAA\textsuperscript{\dag}}), which mitigated the impact of data scarcity.


We also compared the proposed approaches with other state-of-the-art methods in C-SV.
The \textit{Proposed 3/11} model from \cite{singh2024childaugment} was trained using 11 different data augmentation techniques on the out-of-domain dataset VoxCeleb2. We observe that incorporating in-domain child speech CSLU data significantly outperforms using only out-of-domain data augmentation for both intra-year and inter-year verification. Additionally, to assess our proposed approaches against an in-domain model, we compared our models with the \textit{Baseline 3/5} from \cite{singh2024childaugment}, which was trained using 315 hours of MyST child speech data and five data augmentation techniques. Notably, our model fine-tuned with only 24 hours of CSLU-S training data with \textit{++SAA\textsuperscript{*}} method in row 6 of Table \ref{tab:Speaker_and_Feature_adapt_ablation}  achieved an EER of 28.72\% on the IU inter-year evaluation set, outperforming the \textit{Baseline 3/5} model, which achieved an EER of 30.83\%. This highlights that conventional training and augmentation methods alone are insufficient for addressing robustness issues in inter-year C-SV, whereas larger datasets of child speech show more pronounced benefits in intra-year verification. Specifically, we observed that the \textit{Baseline 3/5} model performs notably better on the IU intra-year evaluation set. This is likely because the MyST data (Grades 3 to 5) used for \textit{Baseline 3/5} aligns better with the IU dataset (Grades 1 to 4), while the OGI evaluation sets span a broader age range (Grades 0 to 10).

\begin{table}[t!]
\caption{\textit{Equal Error Rate (EER) comparisons across three test sets are shown. CSLU is an intra-year test set obtained by combining K00 to K10 test sets from CSLU-L evalution split. IU Intra-Year
and IU Inter-Year are combined “Intra-Year” and “Inter-Year” IU test sets from Table \ref{tab:CSLU_IU_Results}. 
\textsuperscript{\textdagger} indicates train data from CSLU-L, and \textsuperscript{*} indicates train data from CSLU-S.}
}


\centering

\resizebox{\columnwidth}{!}{%
\begin{tabular}{lccc}
\hline
\textbf{}                                   & \textit{CSLU} & \textit{IU Intra-Year} & \textit{IU Inter-Year}  \\ \hline  
\textit{Baseline}\textsuperscript{\dag}  & 12.15         & 27.02              & 38.7                      \\ 
\textit{Finetune}\textsuperscript{\dag}  & 11.1          & 19.53              & 35.15                     \\ 
\textit{+FTA}\textsuperscript{\dag}    & 11.58         & 20.30             & 36.97                   \\ 
\textit{++ SAA}\textsuperscript{\dag}   & 8.71    & 19.1               & 34.5                      \\ \hline
\textit{+FTA}\textsuperscript{*}    & \textit{-}         & 12.97             & 31.2                   \\ 
\textit{++ SAA}\textsuperscript{*}   & \textit{-}    & 13.8               & 28.72                      \\ \hline 
\textit{Proposed 3/11} \cite{singh2024childaugment}    & 16.77         & 24.03              & 41.65                     \\ 
\textit{Baseline 3/5}  \cite{singh2024childaugment} & 8.15          & 10.24              & 30.83                     \\ \hline
\end{tabular}}
\label{tab:Speaker_and_Feature_adapt_ablation}
\end{table}

\section{Conclusions}
\label{sec:Conclusion}
In this paper, we propose two approaches to enhance the robustness of children speaker verification (C-SV) systems, focusing on reducing age-related impact. The first approach introduces a Feature Transform Adapter (FTA) to make input acoustic features age-invariant. The second approach enhances data diversity through Synthetic Audio Augmentation (SAA) using HiFi-GAN. We present a new C-SV evaluation dataset with both intra-year and inter-year child speech data. 
We also evaluate our proposed models against the existing C-SV systems and show that in the inter-year evaluation scenario, our best proposed system performs better than the existing C- SV systems.

\section{Acknowledgements}
This work was funded in part by National Science Foundation (NSF) collaborative research grants 1551131 and 1551113.


{\footnotesize
\bibliographystyle{IEEEbib}
\bibliography{strings,refs}
}

\end{document}